**Anisotropic Thermal Conductivity of Exfoliated Black Phosphorus**

*Hyejin Jang[†], Joshua D. Wood[†], Christopher R. Ryder, Mark C. Hersam, and David G. Cahill\**

H. Jang, Prof. D. G. Cahill
   Department of Materials Science and Engineering and Materials Research Laboratory
   University of Illinois
   Urbana, IL 61801, USA
E-mail: d-cahill@illinois.edu

Dr. J. D. Wood, C. R. Ryder, Prof. M. C. Hersam
   Department of Materials Science and Engineering
   Northwestern University
   Evanston, IL 60208, USA

Prof. M. C. Hersam
   Department of Chemistry
   Department of Medicine
   Northwestern University
   Evanston, IL 60208, USA

[†] These authors contributed equally.

**We ascertain the anisotropic thermal conductivity of passivated black phosphorus (BP)**, a reactive 2D nanomaterial with strong in-plane anisotropy. We measure the room temperature thermal conductivity by time-domain thermoreflectance for three crystalline axes of exfoliated BP. The thermal conductivity along the zigzag direction (86 ± 8 W m$^{-1}$ K$^{-1}$) is ~2.5 times higher than that of the armchair direction (34 ± 4 W m$^{-1}$ K$^{-1}$).

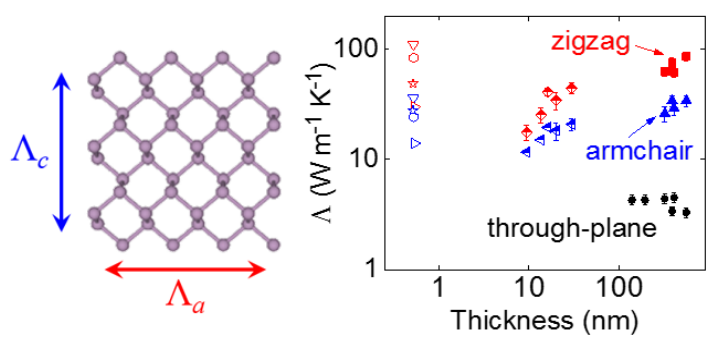

**TOC Figure**

Black phosphorus (BP), a stable phosphorus allotrope at ambient temperature and pressure,[1] is a two-dimensional electronic material with desirable properties for transistor,[2, 3] thermoelectric,[4] and optical sensing[5] applications. Few-layer BP flakes can be exfoliated from bulk crystals due to weak interlayer bonding.[2, 3, 6] In contrast to the planar character of graphite and transition metal dichalcogenides, BP has a puckered, honeycomb structure, leading to heightened chemical reactivity[7] and pronounced in-plane anisotropy. Experimental and theoretical examinations of the electrical,[3, 4, 6] optical,[3, 8] mechanical,[9] and thermal[4, 10, 11, 12] properties reveal distinct anisotropy along BP's two high-symmetry, in-plane directions. These symmetry axes are commonly referred to as the zigzag and armchair directions, with lattice constants of $a$ = 3.314 Å and $c$ = 4.376 Å, respectively.[1]

Understanding an electronic material's thermal conductivity is critical for the thermal management of small-scale devices and for exploring potential thermoelectric applications. Despite extensive electrical characterization of exfoliated BP, experimental measurements of BP's thermal properties are few.[12, 13] First-principles calculations of the anisotropic thermal conductivity of monolayer BP, that is, phosphorene, predict that the thermal conductivity along the zigzag direction is two- or three-fold higher[10, 11] than along the armchair direction; for example, ref. 11 finds 110 and 36 W m$^{-1}$ K$^{-1}$, respectively, in the two directions. Of this, the electronic contribution to the thermal conductivity is markedly small, less than 3 W m$^{-1}$ K$^{-1}$, even at a high carrier concentration of ~$10^{12}$ cm$^{-2}$.[11] Experimentally, Slack found the thermal conductivity of bulk, polycrystalline BP to be ≈10 W m$^{-1}$ K$^{-1}$ at room temperature,[13] but no anisotropic effects were examined. Only mechanically exfoliated BP flakes, with defined symmetry axes, allow an assessment of anisotropic thermal properties in all three high symmetry directions of the crystal. A recent preprint[12] reports the in-plane, anisotropic thermal transport for exfoliated, few-layer BP using micro-Raman spectroscopy; still, the extracted values were much smaller than theoretically predicted for phosphorene. For thinner (<15 nm) BP samples, the measured BP thermal conductivity is modified by phonon scattering from oxidized regions, substrates, and surface imperfections. By contrast, thermal measurements on thicker (>100 nm) BP flakes, especially those protected against ambient oxidation, provide an intrinsic upper bound for the thermal conductivity, as the aforementioned perturbations are less significant. Knowing BP's intrinsic thermal conductivity in its three symmetry axes establishes a needed benchmark for future theoretical and experimental studies of thermal transport in BP, from the bulk to the phosphorene limit.

Herein, we determine the three-dimensional (3D) thermal conductivity tensor of relatively thick, exfoliated BP flakes, which are representative of the intrinsic thermal properties of BP. We passivate the BP flakes with a ~3 nm AlO$_x$ overlayer to protect them from oxidation. The through-plane and in-plane thermal conductivities are measured by conventional time-domain thermoreflectance (TDTR) and beam-offset TDTR. We anticipate that these data will serve as a baseline for thermal management in BP applications and instruct other fundamental phosphorene and few-layer BP studies.

We mechanically exfoliate BP flakes (HQ Graphene) onto Si(100) with a SiO$_x$ native oxide; these flakes are then encapsulated with an AlO$_x$ passivation layer, as detailed previously.[7] We ascertain all flake thicknesses after passivation from fitted height profiles measured in atomic force microscopy (AFM) topographs, which range from 55 to 552 nm. We note that the TDTR techniques require thick flakes (>100 nm) to be able to get reasonable thermal conductivity data. **Figure 1A** and **1B** shows representative optical images of 404 nm and 319 nm thick BP flakes, respectively. As BP is chemically reactive,[7, 14] the ~2.6 nm AlO$_x$ overlayer protects the thick BP flakes from ambient oxidation and intercalation for up to one month (as demonstrated in Figure S1). Before performing TDTR measurements, we coat the passivated BP flakes with a NbV optical transducer. We note that TDTR measurements are not carried out on the flake of Figure S1.

We then orient the BP flakes in their respective zigzag and armchair crystallographic orientations by optical microscope images and angle-resolved polarized Raman spectroscopy. The orientation of a BP flake is adjusted by mounting on the rotating stage, and the deviation from the crystalline axes is expected to be less than 10°. Liu *et al.*[10] determined that the thermal conductivity in BP varies quadratically with the in-plane chiral angle. From their work, we find that a 10° angular deviation gives a ~4% change in the measured thermal conductivity. As this 4% change is less than our other uncertainties, we assert that our angular alignment strategy is satisfactory for our subsequent TDTR measurements. BP's zigzag and armchair axes are indicated schematically in Figure 1C and 1D. We extract a through-plane thermal conductivity by conventional pump-probe TDTR[15] using a frequency-modulated (*f*) pump and a time-delayed probe beam with wavelengths near 785 nm. We determine thermal conductivities in the zigzag and armchair axes of BP by our newly developed beam-offset approach for materials without in-plane symmetry.[16, 17] For the through-plane direction, we employ picosecond acoustics to measure the sound velocity. Additional experimental details are detailed in the Experimental Section.

BP has $D_{2h}^{18}$ point group symmetry and six Raman-active vibration modes.[18, 19] For incident and scattered light propagating normal to the planes of the flakes, the selection rules permit observation of three modes[19] at ~363 cm$^{-1}$ ($A_g^1$), ~440 cm$^{-1}$ ($B_{2g}$), and ~467 cm$^{-1}$ ($A_g^2$). The $A_g^1$, $B_{2g}$, and $A_g^2$ intensities depend on the polarization direction in the *a-c* plane and can be used to determine the crystal orientation.[19] Further, we note that the Raman intensity of the Si TO phonon (~521 cm$^{-1}$) varies with sample orientation, as both the intrinsic Si Raman tensor and BP's optical constants are angle-dependent.

To orient our encapsulated BP samples prior to thermal measurement, we take angle-resolved, parallel polarized Raman spectra, as shown in **Figure 2** and Figure S2. We subtract polynomial baselines from the spectra, Lorentzian fit them, and then normalize their peak intensity values. From Wu *et al.*, the $B_{2g}$ peak intensity is maximum when the scattered polarization is halfway between the zigzag and armchair direction.[19] Conversely, the $A_g^2$ peak intensity is maximum when the polarization is along the armchair direction. In turn, the BP flake of Figure 2 has its armchair axis at 90° relative to the Raman laser polarization axis. We observe no differences for Raman spectra taken on the same axis of an aligned BP flake. This suggests that our exfoliated BP flakes are crystalline and grain boundary free. After orienting the samples, we cap the passivated BP flakes with sputtered NbV. Subsequently, our samples have at least five chemically distinct layers in the following structure (from the top): NbV/AlO$_x$/BP/SiO$_x$/Si(100). The thicknesses of the AlO$_x$ layer and the SiO$_x$ native oxide layer are thin, ~3 nm and ~2 nm, respectively, as measured by ellipsometry; thus, we treat these layers as part of the NbV/BP and BP/Si interface conductances. In our conventional TDTR measurements, we use a relatively large laser spot size, $w \approx 5$ μm, where $w$ is the 1/$e^2$ intensity radius, thereby minimizing the data's sensitivity to the in-plane thermal conductivity.

For through-plane measurements, we inspect BP flakes with thicknesses (here, $h$) between 138 and 552 nm and a diameter > 20 μm. In addition to the typical high modulation frequency of 9.1–9.8 MHz (see Supporting Information), we also use a low frequency ($f$ = 1.1–1.6 MHz) to extend the thermal penetration depth of the laser. Here, the thermal penetration depth is $\sqrt{\Lambda/\pi C f}$, where $\Lambda$, $C$ and $f$ are the thermal conductivity, volumetric heat capacity, and pump modulation frequency, respectively. By combining data at the two modulation frequencies, we can determine the three free parameters in our thermal model: i) the thermal conductance $G_1$ of the NbV/BP interface; ii) the through-thickness $\Lambda_b$ of BP; and, iii) the thermal conductance $G_2$ of the BP/Si interface. We contend that the largest uncertainty source in these experiments is the thickness of the NbV transducer; we measure its thickness using picosecond acoustics and a speed of sound[17] of 5.4 nm ps$^{-1}$. For the volumetric heat capacity, we set $C = 1.87$ J K$^{-1}$ cm$^{-3}$ for the BP layer.[20] Also, for highly anisotropic materials, the anisotropy ratio of an in-plane thermal conductivity to a through-plane conductivity should be included in the thermal model.[16, 17] Therefore, we perform the through-plane analysis iteratively with the following in-plane analysis.

We measure the in-plane thermal conductivities, $\Lambda_a$ and $\Lambda_c$, for thick BP flakes ($h >$ 300 nm), by the beam-offset method.[17] We use a relatively small spot size of $w \approx 2.5$ μm and low modulation frequency ($f = 1.1–1.6$ MHz) to optimize sensitivity to $\Lambda_a$ and $\Lambda_c$. In the beam-offset method, data is collected as the pump beam is translated across the sample surface relative to the probe. The full-width-at-half-maximum (FWHM) of the out-of-phase thermoreflectance signal ($V_{out}$) measured at negative time delay is a robust parameter for determining the in-plane thermal conductivity along the direction of the beam offset.[17] Additionally, the width of the in-phase signal ($V_{in}$) at small positive time delay and high modulation frequency provides a direct measurement of $w$. We remark that the BP flakes thicker than 300 nm are investigated in the beam-offset measurements, due to the limited sensitivity of the thinner flakes on their in-plane thermal conductivities. The FWHM values of such flakes are largely determined by the other experimental parameters, such as the beam spot size ($w$) and the thermal properties of the metal transducer.

Rapid heating of the metal transducer by the pump optical pulses generates a longitudinal acoustic pulse. Reflections of this acoustic pulse that return to the surface, i.e., acoustic echoes, can be used to measure sound velocities if the layer thickness is known. Similarly, acoustic echoes can be used to measure the layer thickness if the sound velocity is known. The sign of the echo depends on the acoustic impedance, $Z = \rho v$, where $\rho$ is the mass density and $v$ the sound velocity of the two materials on the opposing sides of an interface. If the medium underneath has a smaller $Z$, then the acoustic echo undergoes a $\pi$-phase shift.[21] Picosecond acoustics data for a 55 nm thick BP sample are compared to those for the substrate in **Figure 3A**. The first upward echo at 28 ps that repeats periodically is a combination of the reflections from the NbV/AlO$_x$ and AlO$_x$/BP interfaces having a similarly large contrast in $Z$ (see Table S1). When we repeat the measurement on a region of the substrate adjacent to the BP sample, the first peak appears ≈0.5 ps earlier than in the flake, which is reflected predominantly from the NbV/AlO$_x$ interface. This is due to the larger $Z$ ratio of the NbV/AlO$_x$ interface, as compared to that of the AlO$_x$/Si interface. The native oxide of Si, SiO$_x$, generally has a weak effect on picosecond acoustics.[21] The echo at 50 ps is the reflection at the BP/substrate interface. Although the acoustic impedance of BP is smaller than Si, we do not expect an intimately bonded interface between BP and Si, as there is finite roughness for both the exfoliated BP surface and Si substrate. The speed of sound of BP from the measurements of the BP flakes is 4.76 ± 0.16 nm ps$^{-1}$, without showing thickness-dependent behavior (see Figure S3), in excellent agreement with previously reported values[22] for single-crystal BP (namely, 4.5 to 5.1 nm ps$^{-1}$). We also find the elastic constant $C_{33}$ to be $C_{33} = \rho v^2 = 61 \pm 4$ GPa using the bulk BP density[13] of 2.7 g cm$^{-3}$.

**Figure 3B** shows example TDTR data measured at $f$ = 9.1 MHz and fits to the thermal model. The through-plane thermal conductivity is repeatedly analyzed using the anisotropy ratio of the in-plane thermal conductivity to the through-plane one, where the geometric mean of the two in-plane thermal conductivities is used. The best-fitted, through-plane thermal conductivity is $\Lambda_b = 4.0 \pm 0.5$ W m$^{-1}$ K$^{-1}$, and, based on the experimental error, it is insensitive to flake thickness. $\Lambda_b$ of BP is intermediate between MoS$_2$ (2.0 W m$^{-1}$ K$^{-1}$)[23] and graphite (5.4 W m$^{-1}$ K$^{-1}$)[16]. We attribute the low thermal conductivity to the weak, anharmonic van der Waals bonding between the BP layers and the strong anisotropy in the phonon dispersions. Dispersion anisotropy will reduce the thermal conductivity through phonon focusing effects, ultimately reducing the component of the average group velocity in the through-plane direction.[24]

The interface conductance between the NbV and the BP flake is $G_1 = 38 \pm 4$ MW m$^{-2}$ K$^{-1}$, which is relatively low (see Figure S4). While this low interface conductance occurs partly from the low thermal conductivity AlO$_x$ overlayer, it also has been observed for the interface between a metal and a highly anisotropic layered material like graphite[16, 25] and MoS$_2$[23]. The other interface at the bottom of the BP flake, $G_2$, is also estimated low, $G_2 = 60 \pm 10$ MW m$^{-2}$ K$^{-1}$ (Figure S4). This is in accordance with the rough interface characteristics expected from the picosecond acoustics, and also similar to the thermal conductance of a transfer-printed metal film on a substrate.[26]

Beam-offset TDTR measurements for a 384 nm thick BP flake at 1.6 MHz are given in **Figure 4**. We fit Gaussian curves to the TDTR data, allowing us to compare spectra quantitatively by FWHM analysis. FWHM of $V_{out}$ in the zigzag direction, 6.06 ± 0.04 μm, is larger than that of the armchair direction, 5.32 ± 0.03 μm, and both are significantly larger than the width of the laser beam profile, 4.17 ± 0.06 μm. Although these FWHM data can be analyzed using a thermal model[17] that considers the 3D thermal conductivity tensor for anisotropic BP, this 3D model requires multi-dimensional integrations that need significant computational time.

To expedite analysis, we first constrain the range of possible thermal conductivities using a 2D anisotropy model that assumes in-plane symmetry,[16] but we allow the in-plane thermal conductivity to vary depending on the beam-offset direction. Analysis based on the 2D model tends to exaggerate the in-plane anisotropy, due to the incorrect assumption that the in-plane temperature fields are isotropic. For our BP data, the difference between fits to the 2D and 3D models are ≈10%; thus, the 2D model can be used as a useful tool for constraining the values in the thermal conductivities prior to the 3D model's use. All the reported thermal conductivities are derived from the TDTR data using the full 3D model.

To aid in the experimental design and evaluate the propagation of errors, we analyze the sensitivity of the beam-offset FWHM to various experimental parameters.[23] We define the sensitivity parameter for the FWHM as

$$\Sigma(\alpha) = \frac{\partial[\ln(FWHM)]}{\partial[\ln(\alpha)]} \tag{1}.$$

The sensitivities to the experimental parameters over a range of the in-plane thermal conductivities calculated by the 2D anisotropy model are plotted in **Figure 5**. The parameter $h$ is the thickness of the BP flake. Although the model is most sensitive to the beam size, $\Sigma(w)$, we can determine precisely the beam size $w$ during beam-offset measurements. The uncertainty in $\Lambda_r$ originating from a parameter α can be estimated as $\delta(\alpha) \cdot \Sigma(\alpha)/\Sigma(\Lambda_r)$, where $\delta(\alpha)$ is the fractional error of α. Error propagation from uncertainties in $w$ to uncertainties in the in-plane thermal conductivity produces an overall error of ≈10%. The second largest error arises from the determination of thickness ($h$) of BP flakes from AFM measurement; thickness uncertainty is especially important in the case of higher thermal conductivity. Ultimately, we find a 5% uncertainty in the thickness from ten fitted height profiles from AFM, leading to an uncertainty of ≈3% in the in-plane thermal conductivity. The sensitivity to the in-plane thermal conductivity, $\Sigma(\Lambda_r)$, is maximized when the FWHM of $V_{out}$ is larger but comparable to the beam width (see Figure 3B). The $\Sigma(\Lambda_r)$ is maximum at approximately $\Lambda_r = 30$ W m$^{-1}$ K$^{-1}$ for $w = 1$ μm; $w = 2.5$ μm is a better choice for $50 < \Lambda_r < 100$ W m$^{-1}$ K$^{-1}$.

We summarize the thermal conductivities of the BP flakes in **Figure 6**, comparing our results against previous theoretical and experimental studies.[10, 11, 12] In Figure 6, we note that our presented beam-offset TDTR measurements are for thicker ($h > 300$ nm) BP flakes, versus the less sensitive conventional TDTR measurements ($h > 100$ nm) for through-plane BP thermal conductivity determinations. The fitted thermal conductivities are 62–86 W m$^{-1}$ K$^{-1}$ for the zigzag direction and 26–34 W m$^{-1}$ K$^{-1}$ for the armchair direction. Variance in the determined values stems from experimental uncertainty and marginal BP oxidation from changing relative humidity in sample preparation (see Experimental Section). While strongly anisotropic in the plane, the thermal conductivities do not show a statistically significant trend over the range of flake thicknesses investigated. Anisotropic, in-plane thermal conductivities of BP have been predicted for single-layer BP (i.e., phosphorene) from first principles calculations for the phonon dispersions and three-phonon scattering rates.[10, 11] Therein, the phosphorene thickness is set to the inter-layer ($b$ axis) separation of 5.25 Å.[11] Additionally, in a preprint, Luo *et al.* reported[12] the thermal conductivity data for few-layer BP, where they reconciled the thickness dependence of the thermal conductivity through a boundary scattering term to the intrinsic three-phonon scattering rate. Due to the large thickness of the BP flakes we studied in our work, surface scattering should be less significant in our experiments than in those of ref. 12. Further, unlike ref. 12, our BP flakes are encapsulated by both $AlO_x$ and NbV, making ambient oxidation[7] or photodegradation[27] less significant, giving us an intrinsic measurement of BP's anisotropic thermal conductivities.

We briefly note our thermal conductivity results for ambient oxidized BP samples. One BP sample ($h$ > 600 nm thick) was coated with NbV three months after $AlO_x$ encapsulation and measured two weeks after NbV deposition. This sample exhibits much lower in-plane thermal conductivities, namely, 39 ± 4 W m$^{-1}$ K$^{-1}$ and 17 ± 3 W m$^{-1}$ K$^{-1}$ for the zigzag and the armchair directions, respectively. Further it has a lower interface conductance $G_1 \approx$ 30 MW m$^{-2}$ K$^{-1}$, while the through-plane thermal conductivity remains similar, namely, 3.4 ± 0.4 W m$^{-1}$ K$^{-1}$. Comparatively, the samples represented in Figure 6 underwent NbV deposition within one week after encapsulation by $AlO_x$ and were measured within two months of NbV capping. The $AlO_x$ passivation layer is known to protect BP flakes for approximately one month (Figure S1 and ref. 7). It is probable that the NbV deposition can extend the period of sample stability significantly. That notwithstanding, after 5 months, we observe BP degradation from under both the NbV and $AlO_x$ encapsulants (Figure S5). Therefore, while our passivation strategy does not provide indefinite suppression of BP oxidation, it does slow the degradation kinetics enough to allow BP's intrinsic thermal properties to be probed. It is also apparent that prolonged exposure to $O_2$ saturated $H_2O$ vapor prior to NbV deposition promotes sample degradation, affecting the in-plane thermal conductivity but not the through-thickness thermal conductivity. Stronger phonon scattering is reasonable to expect for chemically modified BP, thereby lowering in-plane thermal conductivity. We finally remark that our oxidized in-plane thermal conductivities are similar to those reported in ref. 12. In Luo *et al.*, the BP samples underwent ambient flake transfer, laser irradiation, and solvent exposure for 1–12 hrs. If oxidizing species (e.g., $H_2O$, $O_2$) are not carefully removed during each processing step, then the BP flakes can oxidize,[7] potentially leading to significant changes in the samples' thermal conductivity. Our passivating AlO$_x$ overlayer is designed to protect the sample from oxidation, thereby bounding BP's intrinsic thermal properties.

To understand the intrinsic thickness-dependent behavior of BP, we refer to prior work on other 2D materials: the in-plane thermal conductivity of suspended, few-layer graphene is reported to decrease as additional layers are added,[28] while that of the MoS$_2$ appears to be less sensitive to the thickness.[23] The major difference between the two materials is attributed to the contribution of the out-of-plane acoustic (ZA) mode for the in-plane thermal conductivity[11]: As monolayer graphene is a flat atomic layer, the ZA mode contribution dominates.[11] However, its contribution is greatly reduced in bilayer graphene due to increased inter-layer scattering.[28] By contrast, this dramatic change is not expected in MoS$_2$, as its monolayer is composed of three-atomic planes,[23] where the contribution of the ZA mode is relatively low.[11] The calculated thermal conductivity contributions of different acoustic phonon modes of BP are similar to those in MoS$_2$, especially in the zigzag direction,[11] as expected from the non-planar and puckered structure of BP. Hence, we expect BP's thermal transport to be similar to MoS$_2$. One thing to note is that the flake environment can negatively affect the in-plane thermal conductivity values.[23, 28, 29] Such decreases arise from the contact with a dissimilar support material, the existence of native oxides, or polymer residues at the surface. These environmental factors have been seen for supported graphene[28] or surface disordered, few-layer MoS$_2$.[23, 29] In turn, both of these systems have significantly lower in-plane thermal conductivities than in the bulk.

In summary, we have measured the anisotropic thermal transport properties of passivated, mechanically exfoliated BP flakes. We have probed BP flakes of 138 to 552 nm thickness at room temperature using conventional TDTR and the beam-offset method. The highest in-plane thermal conductivities are 86 ± 8 W m$^{-1}$ K$^{-1}$ and 34 ± 4 W m$^{-1}$ K$^{-1}$ along the zigzag and the armchair directions, respectively. The through-plane thermal conductivity is 4.0 ± 0.5 W m$^{-1}$ K$^{-1}$. The speed of sound in the through-plane direction for BP flakes is estimated to be 4.76 ± 0.16 nm ps$^{-1}$, and the resulting elastic constant, $C_{33}$, is 61 ± 4 GPa. Our in-plane thermal conductivity values are notably high and possess three-fold in-plane anisotropy. These results indicate that the BP is pristine and unoxidized, due to the AlO$_x$ passivation layer that mitigates ambient deterioration. Understanding the significant in-plane thermal anisotropy of BP will help address thermal management for BP and provide opportunities in areas such as in-plane BP thermoelectrics.

**Experimental Section**

*Sample Preparation:* BP flakes were exfoliated in ambient conditions from a bulk crystal (HQ Graphene) onto phosphorus-doped Si(100) wafers (resistivity ≈ 0.3 Ω·cm, University Wafers) with a native oxide ($SiO_x$) thickness of 1.9 ± 0.1 nm as determined by spectroscopic ellipsometry (J.A. Woollam M-2000). The wafers were cleaned with acetone and isopropanol prior to exfoliation. To mitigate BP ambient oxidation,[1, 2] the BP flakes exfoliated onto $SiO_x$/Si were immediately placed in an atomic layer deposition (ALD) reactor (Cambridge NanoTech) and pumped down. The ~319 nm and ~404 nm flakes were exposed to ambient conditions for < 1 min at a relative humidity (RH) of RH ≈ 25%, while the ~384 nm and ~552 nm flakes were exfoliated in < 1 min at RH ≈ 45%.

*Black Phosphorus Passivation and Ellipsometry:* To passivate the BP flakes against ambient oxidation, an alumina ($AlO_x$) overlayer was deposited on the BP/$SiO_x$/Si(100) samples at 50 °C.[2] The passivation layer was grown in two steps. First, adventitious $H_2O$ and $O_2$ on the exfoliated BP flakes were scavenged using 5 cycles of trimethylaluminum (TMA) at 50 °C. Second, the remaining $AlO_x$ was fabricated by using 30 cycles of TMA (0.015 s pulse) followed by anoxic $H_2O$ (0.015 s pulse) at 50 °C. Using Cauchy models on spectroscopic ellipsometry measurements of $AlO_x$/$SiO_x$/Si witnesses, the $AlO_x$ thickness was calculated to be 2.6 ± 0.5 nm.

*Atomic Force Microscopy:* Topographic height images were taken with an Asylum Cypher atomic force microscope (AFM) in normal tapping mode (~320 kHz cantilevers). All images were acquired using the repulsive phase regime with at least 512 samples per line at a scan rate of 1.2 Hz or lower.

*Angle-Resolved Raman Spectroscopy:* BP angle-resolved Raman spectra were taken using a Nanophoton Raman-11 spectrometer with a parallel polarization for the incident and backscattered light. The sample was then rotated about the sample normal in the *a-c* plane. The 532 nm laser power was less than 1 mW and focused by a 20× objective lens; the exposure time for each orientation of the BP flake was 20 s. In addition to the BP Raman peaks, we observed the Si substrate Raman line at 521 $cm^{-1}$.

*Time Domain Thermoreflectance (TDTR):* We measure the thermal conductivities by TDTR, a pump-probe optical technique that uses an ultrafast laser oscillator as the light source. A mode-locked Ti : sapphire laser generates a series of optical pulses with wavelengths near 785 nm at a repetition rate of 74.9–80 MHz. The laser output is split into two orthogonally-polarized beams, the pump and the probe. We use sharp-edged optical filters to spectrally separate the pump and probe.[3] The pump beam heats the surface of the sample and the time-delayed probe beam interrogates the temperature changes at the surface of the sample through the temperature dependence of the optical reflectivity. The pump beam is chopped by an electro-optic modulator at $f = 9.1–9.8$ MHz or $f = 1.1–1.6$ MHz; a photodiode connected to an RF lock-in amplifier detects the changes in the reflected probe intensity that are synchronous with the pump modulation. The sample surface is coated by a 70-nm thick NbV alloy thin film by magnetron sputtering before measurements. We use NbV as the optical transducer, instead of the more common choice of Al, to minimize lateral heat spreading and improve the sensitivity of the beam-offset method to the in-plane thermal conductivity.

**Acknowledgements**
H. J. and J. D. W. contributed equally to this work. Thermal conductivity measurements and analysis at University of Illinois were supported by NSF grant no. EFRI-1433467. Sample preparation and characterization at Northwestern University was supported by the Materials Research Science and Engineering Center (MRSEC) of Northwestern University (NSF DMR-1121262) and the Office of Naval Research (ONR N00014-14-1-0669). The Raman instrumentation at Northwestern was funded by the Argonne-Northwestern Solar Energy Research (ANSER) Energy Frontier Research Center (DOE DE-SC0001059). The authors kindly acknowledge productive discussions with S. A. Wells.

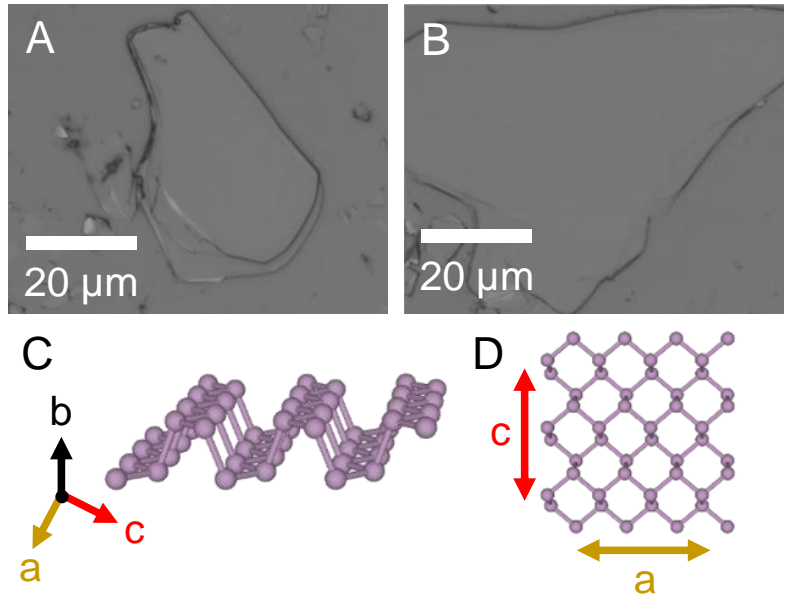

**Figure 1.** Geometry and crystal structure of AlO$_x$ passivated black phosphorus (BP) samples on SiO$_x$/Si(100). **(A, B)** Optical images of passivated BP flakes exfoliated on Si(100). Oblique **(C)** and *a-c* plane **(D)** schematics of the BP lattice structure. Armchair (*c* axis, red) and zigzag (*a* axis, brown) symmetry directions are indicated by the colored arrows.

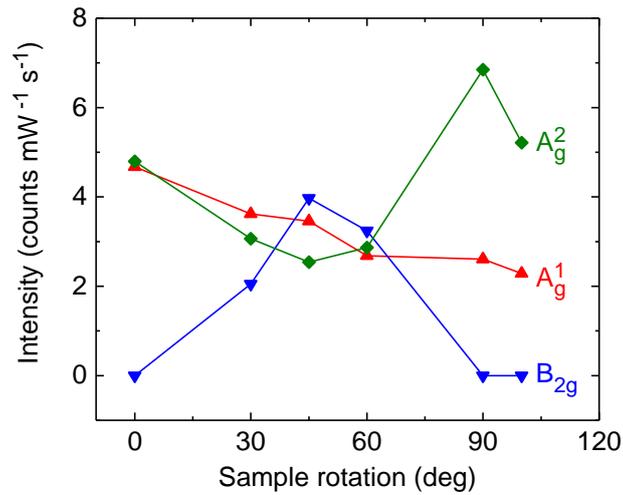

**Figure 2.** Normalized intensities of the three Raman peaks of a BP flake as a function of sample rotation angle with parallel polarization of the incident and backscattered light. The BP flake was rotated in the *a-c* plane with respect to the polarization of the light. The rotation angles of 0° and 90° correspond to the zigzag and the armchair directions, respectively.

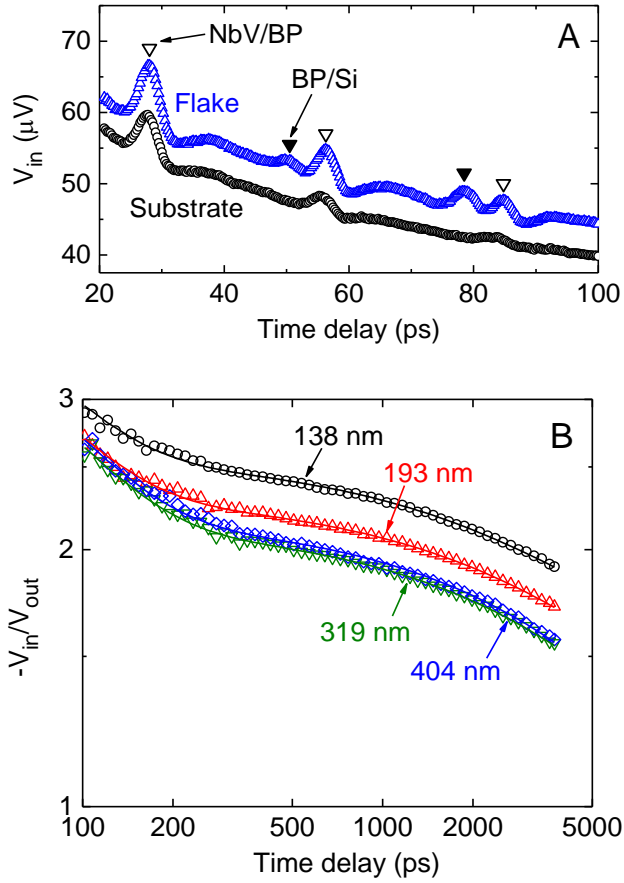

**Figure 3.** Through-plane measurements of the BP flakes coated with the NbV film. **(A)** Picosecond acoustics data for a 55 nm thick BP flake. **(B)** TDTR data for the BP flakes of 138 to 404 nm thicknesses. The open symbols are the measurement data at 9.1 MHz and the solid lines are the best-fits to a thermal model used to determine the through-plane thermal conductivity.

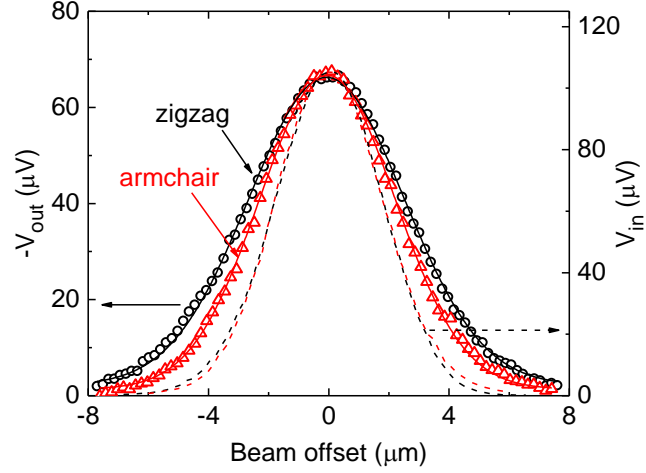

**Figure 4.** In-plane measurements (open symbol) and the Gaussian fitting curves (solid line) of the 384 nm thick BP flake for the zigzag (red) and the armchair (black) directions at low modulation frequency of 1.6 MHz and negative time delay of −50 ps. The dashed lines show the laser beam profiles for each direction determined from the in-phase signal at 9.1 MHz. The $1/e^2$ radius of the pump and probe laser beams is ≈ 2.5 μm.

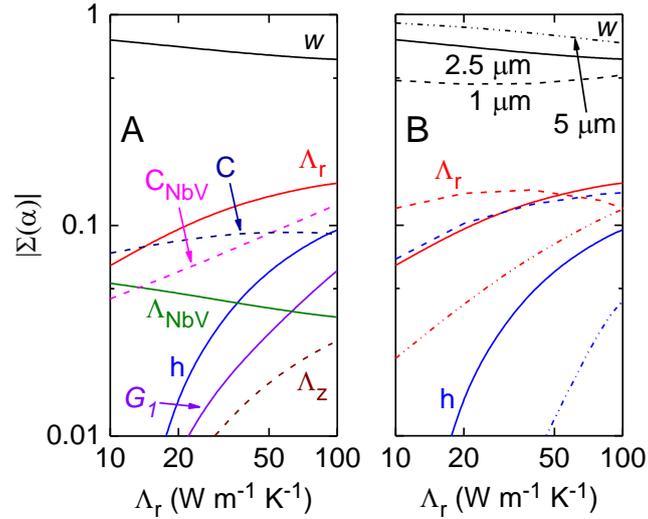

**Figure 5.** Sensitivity of FWHM of the out-of-phase thermoreflectance signal ($V_{out}$) to the experimental parameters for the 552 nm thick BP flake, calculated by the 2D anisotropy model at 1.6 MHz and time delay of −50 ps. **(A)** The sensitivities at the beam size of $w ≈ 2.5$ μm. Dashed lines are sensitivities having negative values. **(B)** The sensitivities of the major parameters at the different beam radii: 1 μm (dash), 2.5 μm (solid), and 5 μm (dash-dot).

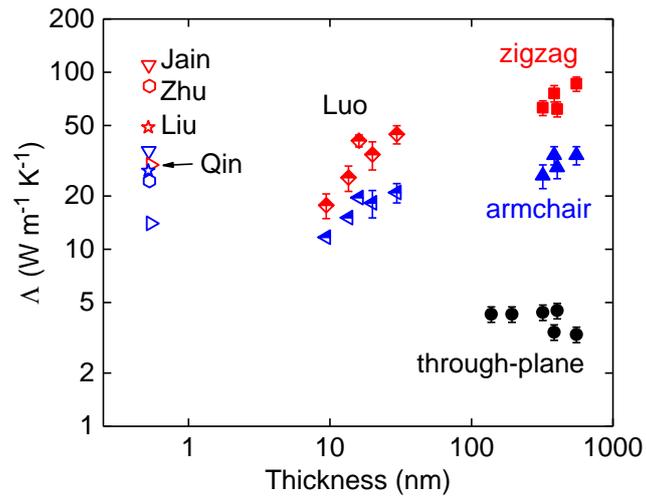

**Figure 6.** Thermal conductivities of BP for the zigzag (red), the armchair (blue), and the through-plane direction (black), plotted in comparison to the reported values as a function of thickness of BP. The monolayer properties are predicted by first-principles calculations (open symbols) and the measured data are reported for few-layered BP (half-filled symbols).

Supporting Information

**Anisotropic Thermal Conductivity of Exfoliated Black Phosphorus**

*Hyejin Jang†, Joshua D. Wood†, Christopher R. Ryder, Mark C. Hersam, and David G. Cahill\**

*E-mail: d-cahill@illinois.edu; † These authors contributed equally.

**Contents:**

- **Figure S1: AlO$_x$ encapsulated black phosphorus (BP) after 1 month in ambient**
- **Figure S2: Angle-dependent Raman point spectra of BP**
- **Figure S3: Longitudinal speed of sound of BP in the through-plane direction**
- **Figure S4: Thermal conductances for the NbV/BP interface ($G_1$) and the BP/Si interface ($G_2$)**
- **Figure S5: BP topography after NbV encapsulation and extended oxidation**
- **Table S1: Acoustic impedances of the material layers in TDTR specimens**

**Supporting Figures**

**Figure S1: AlO$_x$ encapsulated black phosphorus (BP) after 1 month in ambient. (A, B)** Atomic force microscopy (AFM) height images of a AlO$_x$/BP on SiO$_x$/Si(100) after 50 °C AlO$_x$ passivation. **(C, D)** AFM height images of the AlO$_x$ encapsulated flake in **(A, B)** after 1 month in ambient conditions. Sample was stored in the dark at a relative humidity of *ca.* 30%. After ambient exposure, the flake thickness increases from 66.0 ± 2.8 nm to 72.1 ± 0.5 nm, suggesting the intercalation of ambient species. Regardless, typical BP morphological degradation[1] is absent, indicating slowed BP oxidation by the ~2.6 nm AlO$_x$ overlayer. Oxidation can be slowed further by our NbV encapsulation, which is used as a transducer for time-domain thermoreflectance (TDTR) measurements. However, the ~75 nm thick NbV film can obscure our ability to observe subtle, oxidation-induced changes in BP morphology. The flake presented here was never encapsulated with NbV, preventing TDTR measurement.

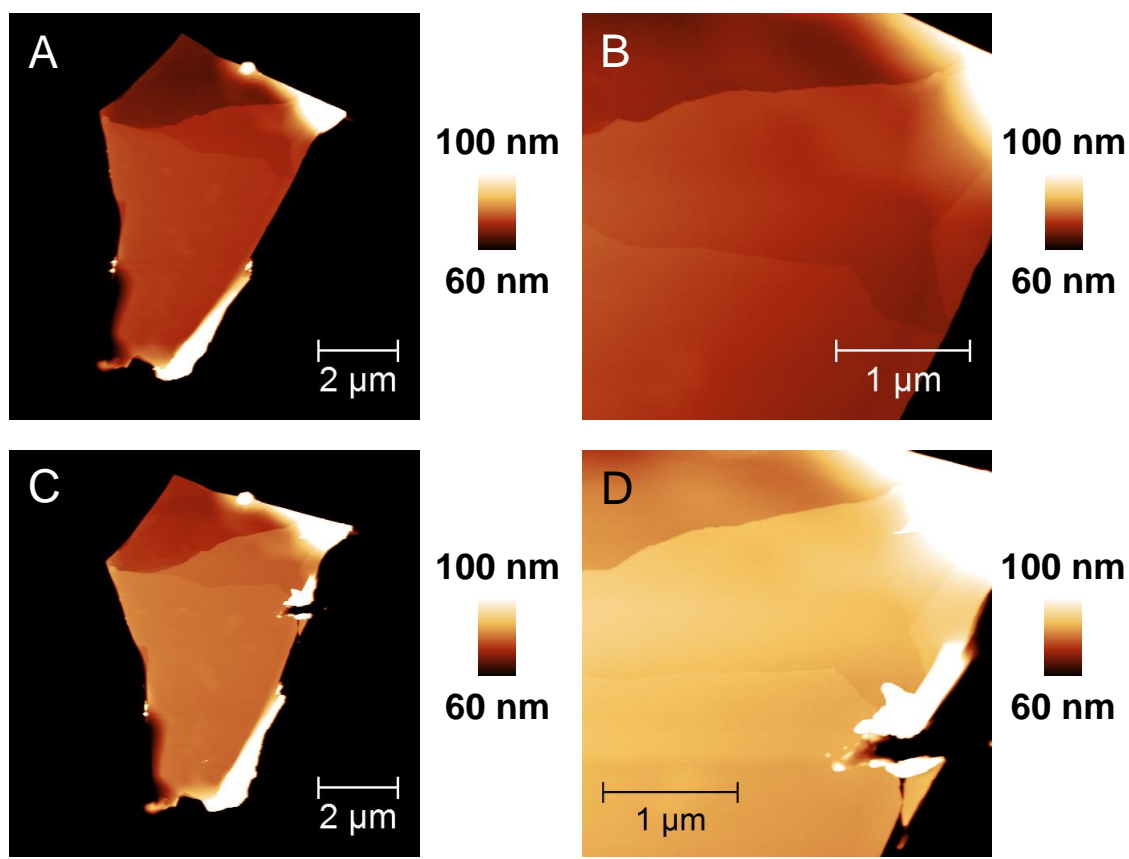

**Figure S2: Angle-dependent Raman point spectra of BP.** Spectra taken with parallel polarization of the incident and backscattered light, and spectra normalized and offset for clarity. The BP flake was rotated with respect to the polarization of the light in the *a-c* plane. The y-axis is the spectral CCD readout, divided by the incident laser power in mW and the integration time in s.

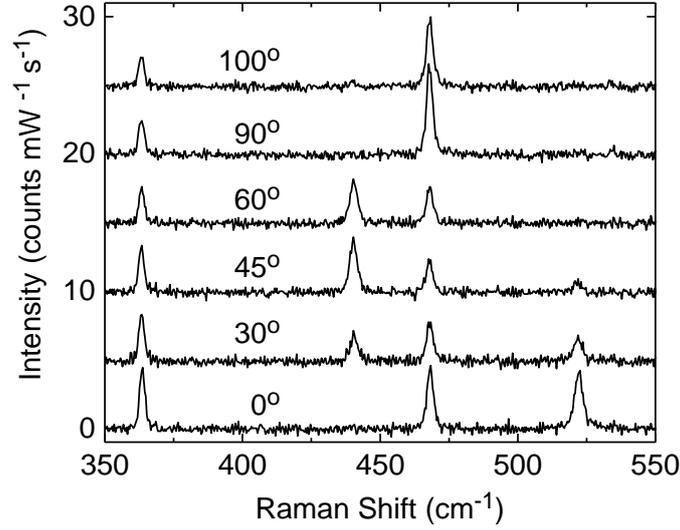

**Figure S3: Longitudinal speed of sound of BP in through-plane direction (*b*-axis).** The speed of sound is determined from picosecond acoustics data of the BP flakes having a thickness 55 to 404 nm. The speed of sound is insensitive to flake thickness, with an average value of $4.76 \pm 0.16$ nm ps$^{-1}$.

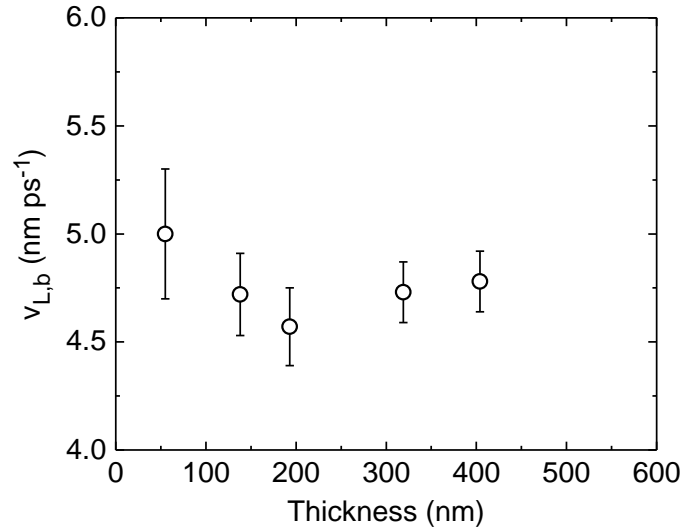

**Figure S4: Thermal conductances for the NbV/BP interface ($G_1$) and the BP/Si interface ($G_2$).** Both $G_1$ and $G_2$ are determined from the through-plane TDTR measurements by using the two different modulation frequencies.

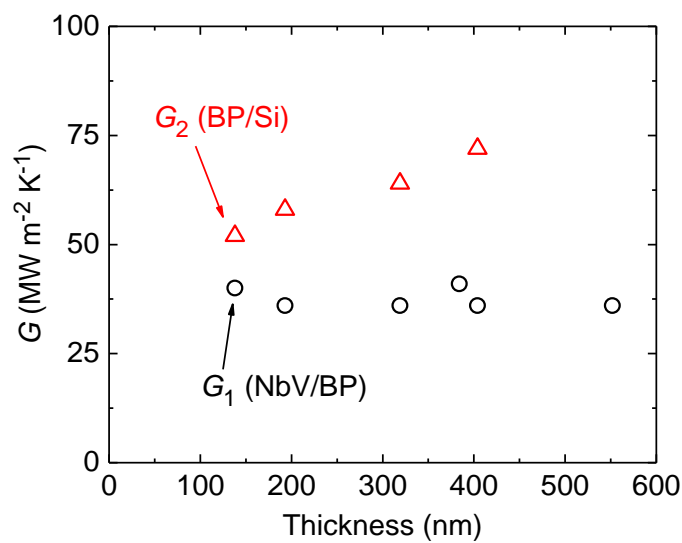

**Figure S5: BP topography after NbV encapsulation and extended oxidation.** AFM height images of a 552 nm thick flake **(A)** before NbV encapsulation and **(B)** after NbV capping and 5 months in ambient. This flake was passivated by a ~2.6 nm AlO$_x$ overlayer before NbV encapsulation. Arrow denotes the same location on the flake. After 5 months of ambient aging, the NbV/AlO$_x$ capped BP flake has dendritic features suggestive of intercalated species and oxidation.

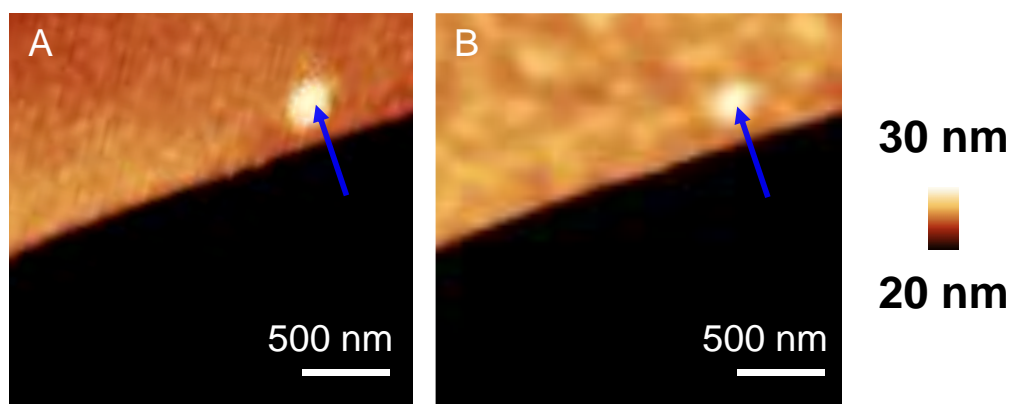

**Table S1: Acoustic impedances of the material layers in TDTR specimens.** $\rho$, *v*, and Z are the density, the speed of sound, and the acoustic impedance of each material, respectively.

| Materials | $\rho$ (g cm$^{-3}$) | *v* (nm ps$^{-1}$) | Z = $\rho v$ |
|---|---|---|---|
| NbV [2] | 7.4 | 5.4 | 40 |
| AlO$_x$ [3] | 2.5 | 8.7 | 22 |
| BP | 2.7 | 4.76 | 13 |
| SiO$_x$ [4] | 2.2 | 5.97 | 13 |
| Si [4] | 2.33 | 8.48 | 20 |